\documentclass[12pt]{iopart}

\usepackage{iopams}

\usepackage{psfig}

\usepackage{amssymb}
\usepackage{amsfonts}
\usepackage{bm}
\usepackage{bbm}

\newcommand{\eqname}[1]{\label{eq:#1}}

\newcommand{\el}{{{\bf e}}}
\newcommand{\EE}{{{\bf E}}}

\newcommand{\kk}{ {\bf k}}

\newcommand{\xx}{ {\bf x}}

\newcommand{\rr}{ {\bf r}}
\newcommand{\qq}{ {\bf q}}

\newcommand{\eq}[1]{(\ref{eq:#1})}

\newcommand{\Psih}{\hat\Psi}

\newcommand{\upd}{d}

\begin{document}

\title[Bragg scattering and spin structure factor]{
Bragg scattering and the spin structure factor of two-component atomic
gases}

\author{I. Carusotto}
\address{CNR-INFM BEC Center and Universit\`a di Trento, 38050 Povo, Italy}
\ead{carusott@science.unitn.it}

\begin{abstract}
Bragg scattering with linearly polarized light can be used to
separately measure the density and the spin structure
factor of a two spin component atomic gas by looking at the dependance
of the scattering intensity on the polarization of the laser fields.
Both stimulated and spontaneous scattering are discussed. Explicit
results for different spin configurations are given.
\end{abstract}



\section{Introduction}

In recent years, a strong activity has been concentrated on the study
of atomic gases with spin degrees of freedom~\cite{SpinGeneral}.
In this context, optical diagnostic techniques appear very
promising, as they provide the possibility of selectively addressing the
different spin states by playing with the light frequency, as in the first
observation of vortices in Bose-Einstein
condensates~\cite{VorticesJILA}, or with the polarization degrees of
freedom, as in recent experimental studies of the magnetization in a spin-1 atomic  
cloud~\cite{SpinorStamper,IC-EM}.

Among the many effects under investigation,
the superfluid transition  of 
two spin component Fermi gases close to a Feshbach resonance~\cite{ExpBCS_K,ExpBCS_Li}
has received a particular interest from both the theoretical and the
experimental side since atomic gases are clean systems where it is 
possible to quantitatively test different theories for
strongly-interacting many-fermion systems~\cite{TheoryBCS}. 
One of the key observables for the microscopic characterization of
such a system are the density-density correlation functions. In
particular the opposite spin correlation function $S_{\perp}(\xx,\xx')=\langle
n_+(\xx)\,n_-(\xx')\rangle$ carries information on the pairing effect
which is responsible for superfluidity~\cite{SupercondBooks}.
Recent experiments using molecular spectroscopy techniques have
measured its local value
$S_{\perp}(\xx=\xx')$~\cite{DensityCorrelationsExp}, but no data are yet
available at different points $\rr\neq\rr'$. 

Scattering techniques are widely applied to the study of condensed matter
systems, since they allow for a precise characterization of the
structure and the dynamics down to length scales of the order of the
wavelength used in the experiment. 
Remarkably, the typical microscopic length scale of ultracold atomic
gases is often on the order of the visible wavelength, and this has
allowed the use of stimulated Bragg scattering techniques, e.g. to
measure the structure factors (i.e. the Fourier transform of the
density-density correlation function) in clouds of spinless
bosonic atoms~\cite{Bragg}.

In the present paper, we propose to take advantage of the polarization
degrees of freedom of light in order to perform Bragg scattering experiments
which are able to selectively address the total density and the spin
density of the atomic gas, and then give information on the
same- and opposite-spin density-density correlations.
After a brief review of the light-matter interaction for spin-1/2
atoms, we shall show that a stimulated Bragg scattering
experiment analogous to the one of~\cite{Bragg} is able to measure the 
response function to perturbation 
operators which are a combination of the total and the spin densities with
polarization-dependent weights. The fluctuation-dissipation theorem
then relates the absorbed energy to the structure factor.
A direct measurement of the structure factor, not relying on the
fluctuation-dissipation theorem, can in principle be performed by means of a
spontaneous Bragg scattering experiment using a single incident beam
and a polarization-sensitive detection.

\section{The optical response of spin 1/2 atoms}
\label{sec:general}

Consider a gas of spin $F_g=1/2$ fermionic atoms with an optical
transition at frequency $\omega_{eg}$ to an excited state of spin
$F_e$. 
In the simplest case $F_e=1/2$, the light-matter coupling Hamiltonian
in the rotating-wave approximation can be written in the compact
form~\cite{IC-EM}: 
\begin{equation}
H_{\rm lm}=\frac{d}{\sqrt{2}}\,\int\!\upd\rr\,\sum_{\alpha\beta=\pm}{\bm
  \sigma}^i_{\alpha\beta}\,{\hat
  E}_i(\rr)\,\Psih^{\dagger}_{e,\alpha}(\rr)\,\Psih_{g,\beta}(\rr)+\textrm{H.c.},
  \label{eq:light_matter_coupl}
\end{equation}
in terms of the electric dipole moment $d$ of the optical transition.
Here, ${\bm \sigma}^i$ are the Pauli matrices ($i=\{x,y,z\}$), and
${\hat E}_i$ are the electric field operators; $\Psih_{g,\alpha}$ and
$\Psih_{e,\alpha}$ are the atomic field operators for the Zeeman
sublevels $\alpha,\beta=\pm$ of respectively the ground $g$ and the
excited $e$ states.
In the $F_e=3/2$ case, the structure is analogous, the index $\alpha$
now running over the four Zeeman sublevels $\alpha=3/2, 1/2, -1/2, -3/2$
of the excited state, and the Pauli matrices being replaced by the
ones of the relevant Clebsch-Gordan coefficients.

Assuming that the light interacting with the atoms is far from
resonance with the optical transition, we can adiabatically eliminate
the excited state and write an effective Hamiltonian involving the
ground state only. If no external field is present which breaks the
rotational symmetry, the Zeeman sublevels of 
both the ground and the excited states are 
degenerate, so that the different tensorial components of the light
and matter fields can be separated, as originally done in~\cite{CCT}.
For our $F_g=1/2$ case, the effective Hamiltonian has then the
form~\cite{Texas}:
\begin{equation}
  \label{eq:light_matter_coupl_2}
  H_{\rm eff}=\frac{1}{2}\,\int\!\upd\rr\,
\left( f_n\,\delta_{ij}\,\delta_{\alpha\beta}+
i\,f_s\,\epsilon_{ijk}{\bm \sigma}_{\alpha\beta}^k\right)
{\hat E}^\dagger_i(\rr)\,\Psih^{\dagger}_{g,\alpha}(\rr)
\,\Psih_{g,\beta}(\rr)\,{\hat E}_j(\rr),  
\end{equation}
where repeated indices are meant to be summed over.
The effective coupling constants $f_{n,s}$ depend on the
electric dipole moment $d$ of the optical transition and on the
detuning from the excited state.
For the simplest case of a $F_g=1/2\rightarrow F_e=1/2$ transition,
one has \mbox{$f_n=f_s=|d|^2/\hbar(\omega_L-\omega_{eg})$}, while for a
  $F_g=1/2\rightarrow F_e=3/2$ transition one has
\mbox{$f_n=-2\,f_s=4\,|d|^2/3\hbar(\omega_L-\omega_{eg})$}.
If several transition are involved, their contributions to $f_{n,s}$
have to be algebraically summed.

By means of elementary vectorial analysis, this Hamiltonian
can be rewritten as:
\begin{equation}
H_{\rm eff}=\frac{1}{2}\int\!\upd\rr\,\left[f_n\,{\hat n}(\rr)\,{\hat
    I}(\rr)+f_s\,{\hat {\mathbf 
    S}}(\rr)\cdot{\hat{\mathcal
    B}}(\rr)\right]  
\eqname{Ham_eff}
\end{equation}
in terms of the intensity ${\hat I}(\rr)$ and the pseudo-magnetic field
${\hat {\mathcal B}}(\rr)$ defined as~\cite{Texas}:  
\begin{eqnarray}
{\hat I}(\rr)&=&{\hat {\mathbf
    E}}^\dagger(\rr)\cdot{\hat {\mathbf E}}(\rr) \\
{\hat {\mathcal B}}(\rr)&=&i\,{\hat{\mathbf E}}^\dagger(\rr)\times {\hat
    {\mathbf E}}(\rr).
\end{eqnarray}
Note how the coupling of the light field to the electric dipole of the
far-off resonant optical transition results in the effective potential
\eq{Ham_eff} which not only couples to the atom density ${\hat
  n}(\rr)=\Psih^\dagger_{g,\alpha}(\rr)\,\Psih_{g,\alpha}(\rr)$, but
also to the spin density ${\hat
  S}_i(\rr)={\bm
  \sigma}_{\alpha\beta}^i\,\Psih^{\dagger}_{g,\alpha}(\rr)\,\Psih_{g,\beta}(\rr)$. 
The physical meaning of this coupling can be easily understood at least in some
most significant cases.
For linearly polarized light, the vector product vanishes,
so that the optical potential couples to the total density only.
On the other hand, for a $\sigma_+$-polarized light, the
pseudo-magnetic field is ${\mathcal B}=-I\,{\hat z}$. In the case of a
transition to a $F_e=1/2$ excited state for which $f_n=f_s$, it is
easy to see from \eq{Ham_eff} that the optical potential \eq{Ham_eff} created by
$\sigma_+$-polarized light acts on the $-$ state only.
No optical transition can in fact be driven on atoms in the $+$ state
by $\sigma_+$ light.


\section{Stimulated Bragg scattering}
\label{sec:stimulated}

Consider a stimulated Bragg scattering configuration in which 
a pair of laser beams of wavevectors $\kk_{1,2}$ are sent onto the atomic
cloud with polarization vectors ${\bf E}_{1,2}$. 
If the laser frequencies  $\omega_{1}$ and
$\omega_2=\omega_1-\Delta\omega$ are well detuned from atomic
resonance, spontaneous 
emission can be neglected, and the electric field operators in the
Hamiltonian \eq{Ham_eff} can be replaced by the
classical ${\bf C}$-number amplitudes given by the 
superposition ${\bf E}(\rr)={\bf E}_1(\rr,t)+{\bf E}_2(\rr,t)$ of the
two beams ${\bf E}_{1,2}(\rr,t)={\bf 
  E}^o_{1,2}(\rr)\,e^{i\kk_{1,2}\rr}\,e^{-i\omega_{1,2}t}$.
The spatial dependence of ${\bf E}_{1,2}^o(\rr)$ accounts for the finite
waist of the two beams.

The spin-dependent optical potential acting on the atoms then reads:
\begin{eqnarray}%
H_{\rm
  eff}&=&\frac{1}{2}\,\int\!\upd\rr\,\Big[f_n\,\Big(\EE^{o*}_1(\rr)\cdot\EE^o_1(\rr)+
\EE^{o*}_2(\rr)\cdot\EE^o_2(\rr)\Big)\,{\hat n}(\rr)+ \nonumber \\
&+&i f_s\,\Big(\EE_1^{o*}\times \EE^o_1+\EE_2^{o*}\times \EE^o_2\Big)\cdot
  {\hat {\bf S}}(\rr)+ \nonumber \\
&+&f_n\,\Big(\EE^{o*}_2(\rr)\cdot\EE^o_1(\rr)\,e^{-i(\qq\cdot\rr+\Delta\omega\,t)}+\textrm{h.c.}\Big)\,{\hat
  n}(\rr)+ \nonumber \\
&+&i\,f_s\,\Big(\EE_2^{o*}\times
  \EE^o_1\,e^{-i(\qq\cdot\rr+\Delta\omega\,t)}+\textrm{h.c} \Big)\cdot
  {\hat {\bf S}}(\rr)
\Big].
\eqname{opt_pot}
\end{eqnarray}%
The first two terms describe the optical potential created by
each laser beam separately, and has a negligible effect on the atoms
as long as the waist of the beams is much wider than the atomic cloud.
The third and fourth terms describe the potential due to the
interference pattern of the two beams.
This has a wavevector $\qq=\kk_2-\kk_1$ and moves at speed
$v=-\Delta\omega/q$ in the direction of $\qq$. 

A very interesting case is when both beams have linear polarizations
along the directions $\el_{1,2}$ making a relative angle $\theta$ such
that $\cos\theta=\el_1\cdot \el_2$. Neglecting the effect of the finite 
waist of the beams, the optical potential Hamiltonian \eq{opt_pot} can
be rewritten as:
\begin{eqnarray}%
H_{\rm eff}=\frac{E_2^{o*}\,E_1^o}{2}\,\int\!\upd\rr\,\Big[f_n\,\big(\el_2\cdot\el_1\big)\,{\hat n}(\rr)&+&
i\,f_s\,\big(\el_2\times \el_1\big)\cdot {\hat {\bf
    S}}(\rr)\Big]\,e^{-i(\qq\cdot\rr+\Delta\omega\,t)}+ \nonumber \\
&+&\textrm{H.c.}
\eqname{opt_pot_2}
\end{eqnarray}%
which has the standard form of a linear response Hamiltonian:
\begin{equation}
H_{\rm eff}=V_0\,{\hat A}\,e^{-i\Delta\omega\,t}+\textrm{H.c.},
\eqname{LinResp}
\end{equation}
the perturbation operator ${\hat A}$ being:
\begin{equation}
{\hat A}=f_n\,\cos\theta\,{\hat n}(\qq)+
i\,f_s\,\sin\theta\,{\hat S}_z(\qq),
\eqname{pert_op}
\end{equation}
where 
\begin{eqnarray}
{\hat n}(\qq)&=&\int\!d\rr\,e^{-i\qq\cdot\rr}\,{\hat n}(\rr) \\
{\hat {\bf S}}(\qq)&=&\int\!d\rr\,e^{-i\qq\cdot\rr}\,{\hat {\bf S}}(\rr)
\end{eqnarray}
are the Fourier components of the total and spin densities.

Depending on the relative weight of the coupling to the density and to
 the spin density, different families of elementary
excitations~\cite{SupercondBooks,Yvan,Buchler} are addressed by  the perturbation ${\hat A}$.
The Anderson-Bogoliubov phonon~\cite{AndersonPhonon} couples in fact
 to the density, while pair-breaking excitations couple to the spin density.
Here, one can go from one case to the other simply by varying the
 angle $\theta$ between the linear polarizations $\el_{1,2}$.
If $\el_{1,2}$ are parallel ($\theta=0$), the field is everywhere
 linearly polarized, with a sinusoidal spatial dependance.
This means that the vector product in \eq{opt_pot_2} vanishes and the optical
potential couples to the total density only.
On the other hand, if $\el_{1,2}$ are orthogonal $\theta=\pi/2$
(lin~$\perp$~lin configuration~\cite{Lattice}), the field intensity
$I(\rr)$ is
spatially constant, but the polarization periodically varies in
space as $\sigma_+\rightarrow ({\hat x}+{\hat  y})\rightarrow \sigma_-
\rightarrow ({\hat x}-{\hat  y}) \rightarrow \sigma_+$ giving a
sinusoidally varying pseudo-magnetic field
${\mathcal B}(\rr)$ polarized along the direction perpendicular to
$\el_{1,2}$. In the following we shall take this direction as the
 quantization $z$ axis: in this way, the  optical potential results
 always diagonal in the $\pm$ basis of the atomic states. If a
 different quantization axis is chosen, e.g. orthogonal to
 $\el_1\times \el_2$, the coupling of the pseudo-magnetic field
 ${\mathcal B}(\rr)$ to the spin would result  in off-diagonal terms
 coupling the two different spin states in a Raman-like way.

It is a general result of linear response theory~\cite{LinearResponse}
that the imaginary part 
$\textrm{Im}[\chi_{A^\dagger A}(\Delta\omega)]$ of the {\em response
function} $\chi_{A^\dagger A}$ can be measured from the amount of
energy (or momentum) transferred to the system for a given excitation
sequence of the form \eq{LinResp}.
For a quasi-monochromatic excitation at frequency $\Delta\omega$, the
transferred energy is indeed
proportional to $E_{\rm
  abs}\propto\Delta\omega\,\textrm{Im}[\chi_{A^\dagger
    A}(\Delta\omega)]$. 
For a system at thermal equilibrium at a temperature $T$, the
fluctuation-dissipation theorem~\cite{LinearResponse} then relates 
$\textrm{Im}[\chi_{A^\dagger 
  A}(\Delta\omega)]$ to the corresponding {\em dynamic structure factor}
$S_{A^\dagger A}(\Delta\omega)=
\int\!\upd t\,\langle A^\dagger(t)\,
A(0)\rangle\,e^{i\Delta\omega\, t}$:  
\begin{equation}
  \label{eq:FD}
  \textrm{Im}[\chi_{A^\dagger
      A}(\Delta\omega)]=-\frac{1}{2\hbar}\,S_{A^\dagger A}(\Delta\omega)\,
\left(1-e^{-\hbar\,\Delta\omega/k_B T} \right).
\end{equation}
In our case, $S_{A^\dagger A}$ can be written as:
\begin{equation}%
S_{A^\dagger A}(\Delta\omega)=f_n^2\,\cos^2\theta\,S_n(\qq,\Delta\omega)+f_s^2\,\sin^2\theta\,S_s(\qq,\Delta\omega),
\eqname{S_A}
\end{equation}
in terms of the density and spin dynamical structure factors
$S_{n,s}(\qq,\Delta\omega)$ defined as:
\begin{eqnarray}
S_n(\qq,\Delta\omega)&=&
\int\!\upd t\,\upd \rr\,
\Big\langle
{\hat n}(\rr,t)\,
{\hat n}(0,0)\,
 \Big\rangle\,
e^{-i(\qq\cdot \rr-\Delta\omega\, t)} \\
S_s(\qq,\Delta\omega)&=&
\int\!\upd t\,\upd \rr\,
\Big\langle
{\hat S}_z(\rr,t)\,
{\hat S}_z(0,0)\,
 \Big\rangle\,
e^{-i(\qq\cdot \rr-\Delta\omega\, t)}.
\end{eqnarray}

The contributions of respectively the total and the spin density can
be then isolated by measuring the response function
$\textrm{Im}[\chi_{A^\dagger A}]$ for different values of
the angle $\theta$ between the two polarizations.
Knowledge of the total and spin density structure factors immediately
 gives the same- and opposite-spin density correlations:
\begin{eqnarray}
S_{\parallel,\perp}(\qq,\Delta\omega)&=&
\int\!\upd t\,\upd \rr\,
\Big\langle
{\hat n}_+(\rr,t)\,
{\hat n}_\pm(0,0)\,
 \Big\rangle\,
e^{-i(\qq\cdot \rr-\Delta\omega\,
  t)}= \nonumber \\
&=&\frac{1}{4}\Big[S_n(\qq,\Delta\omega)\pm
  S_s(\qq,\Delta\omega)\Big] 
\end{eqnarray}
Differently from the dynamic structure factor, the low-temperature
 response function $\chi_{A^\dagger A}$ is generally weakly 
 dependent on $T$, and is well approximated by its $T=0$ value.
Under this approximation, by setting $T=0$ in \eq{FD}, it is immediate
 to see that a measure of  $\chi_{A^\dagger A}$ provides a good
 approximation of the most relevant zero temperature value of $S_{A^\dagger
 A}(\Delta\omega)$.
The static structure factor $S_{A^\dagger
 A}=\langle A^\dagger(t) A(t) \rangle$ is obtained
 from the corresponding dynamical one $S_{A^\dagger
 A}(\Delta\omega)$ by frequency integration:
\begin{equation}
S_{n,s}(\qq)=\int\!\frac{d\Delta\omega}{2\pi}\,S_{n,s}(\qq,\Delta\omega)\,e^{-i\Delta\omega\,t}.
\end{equation}

\section{Spontaneous scattering}


So far, we have considered the case of stimulated Bragg scattering,
where two laser beams were sent on the cloud and scattering out of the 
first was stimulated by the presence of the second one.
Now we shall analyse a different configuration, where information on
the spin and density correlations in the atomic gas can be directly
retrieved without invoking the fluctuation-dissipation theorem.
Let us consider a spontaneous Bragg scattering process in which 
a single, linearly polarized, coherent laser beam is sent onto the atomic cloud with a
wavevector $\kk_1$ and amplitude $\EE_1=E_1^o\,\el_1$ and 
the intensity of spontaneously scattered light with wavevector
$\kk_2$, linear polarization $\el_2$, and frequency
$\omega_2=c k_2=\omega_1-\Delta\omega$
is measured as a function of the scattering wavevector
$\qq=\kk_2-\kk_1$ and the polarizations $\el_{1,2}$.

For the sake of simplicity, we start from the case where no external
field is present, so that the light-matter coupling
Hamiltonian \eq{Ham_eff} can be used.
If the laser frequency $\omega_1$ is far enough from the resonance
frequency $\omega_{ge}$, the atomic cloud is optically dilute
and optically thin. Multiple scattering processes can then be neglected.
Within this Born approximation, the (vector, operator-valued)
scattering amplitude can be written as:
\begin{equation}
{\hat \EE}(\kk_2)\simeq \frac{E_1^o}{2}\,{\mathcal P}_{\kk_2}\,\big(f_n
\,{\hat n}(\qq,\Delta\omega)\,\el_1+i\,f_s\,\el_1\times{\hat {\bf
      S}}(\qq,\Delta\omega) \big),
\label{eq:scattered_ampl}
\end{equation} 
where $\kk_2$ is the wavevector of the scattered photon, ${\hat
  n}(\qq)$ and ${\hat {\bf S}}(\qq)$ are the Fourier components of the
  total density and the spin density, and the projector ${\mathcal 
  P}_{\kk_2}$ projects onto the plane orthogonal to 
$\kk_2$ in order to account for the transverse nature of the e.m. field.
This expression has a transparent physical intepretation in terms of a
fluctuating dielectric susceptibility of the atomic gas~\cite{IC-EM}.
The total density corresponds to the scalar, isotropic, component of
the dielectric tensor, while the spin density gives an optical
activity which rotates the polarization plane of the light.
Fluctuations at a wavevector $\qq$ of the susceptibility tensor
scatter light from the incident mode at $\kk_1$ into the one at $\kk_2$.
Note that no nematic term of the form discussed in~\cite{IC-EM} for
spin-1 atoms can be found in the present spin-1/2 case because of
symmetry arguments.

As we are considering atomic clouds whose density profile is
smooth, the spatial Fourier transform of the average total and spin
densities are non-vanishing only at wavevectors $q$, much
smaller than the inverse cloud size $1/R$.
This implies that the coherent scattering amplitude $\big\langle {\hat
  {\bf E}}(\kk_2)\big\rangle$ is not vanishing only in a small cone of
aperture $\lambda/ R$ around the incident beam direction,
corresponding to the standard diffraction on a refractive 
body of size $R$. 
In the following, we shall place ourselves well outside this cone,
where only incoherent scattering occurs.

Interesting information on the density fluctuations in the cloud can
be retrieved from the properties of the incoherently scattered light,
in particular from the dependence of its intensity on the polarization
vectors $\el_{1,2}$.
In practice, the scattered intensity $I_{\rm sc}$ into the polarization state
$\el_2$ can be isolated by making the scattered light at $\kk_2$ to pass
through a linear polarizer before reaching the detector.
As $\el_2$ is by definition orthogonal to $\kk_2$, the intensity of
the scattered light can be written as:
\begin{eqnarray}
I_{\rm sc}&=&\Big\langle \big(\el_2\cdot {\hat {\bf E}}^\dagger(\kk_2)\big) 
\big(\el_2\cdot{\hat {\bf E}}(\kk_2)\big) \Big\rangle= \\
&=&\frac{1}{4}\Big[f_n^2\,\cos^2\theta\,
  S_n(\qq,\Delta\omega)+f_s^2\,\sin^2\theta\,
  S_s(\qq,\Delta\omega)\Big]\,I_{\rm inc}
\eqname{I_2}
\end{eqnarray}
where the quantization axis $z$ for the spin has been again chosen
along the $\el_2\times \el_1$ direction, $I_{\rm inc}=|E_1^o|^2$ is
the incident intensity and $\theta$ is the angle between the linear
polarization vectors  
$\el_{1,2}$.
Note the strict analogy between the result \eq{I_2} for spontaneous
scattering, and formula \eq{S_A} for the stimulated one. 
Although different quantities are experimentally measured in the two cases (the
response function in the stimulated scattering, the structure
factor in the spontaneous one), the
atomic operator involved in the scattering process is the same, as one
can also verify by comparing \eq{opt_pot_2} to the expression for
$\el_2\cdot {\hat {\bf E}}(\kk_2)$ obtained from \eq{scattered_ampl}.
As happens in the stimulated scattering case, the contribution of
the density and spin structure factors to \eq{I_2} can be
isolated simply by repeating the scattering experiment for different
values of the angle $\theta$ between the polarization vector of the
incident beam and the polarization direction of the polarizer used for
analysing the scattered light.

In order to directly access the static structure factor
$S_{n,s}(\qq)$, one has to detect the whole spectrum of the
scattered light (as in X-ray scattering from crystals~\cite{A-M}) or,
better, to perform the experiment using a short pulse 
containing a wide range of frequencies as the incident beam. 
With respect to the stimulated Bragg scattering discussed in the
previous section, the present technique has therefore the advantage of
not requiring the repeated measurement for different detunings
$\Delta\omega$ in order to get the static structure factor
\footnote{Note that the stimulated Bragg scattering experiment
 discussed in sec.\ref{sec:stimulated} can not be used in the short
 pulse regime in order to directly measure the static structure factor.
An extra $\Delta\omega$ factor is in fact present in the proportionality law
 between the absorbed energy $E_{\rm abs}$ and the quantity of interest
 $\textrm{Im}[\chi_{A^\dagger A}(\Delta\omega)]$, so that the total
 absorbed energy in a pulsed experiment is trivially determined by the
 $f$-sum rule~\cite{f-sum}: 
\begin{equation}
\int\!d\omega\,\omega\,\chi_{n,s}(\qq,\omega)=N\frac{\hbar q^2}{2m},
\end{equation}
rather than by the static structure factor.}. 
This is easily understood in physical terms: a scattering experiment
performed with a short pulse compared to the characteristic
time-scale for the atomic motion provides an instantaneous measurement
of the atomic positions, i.e. of the static structure factor.

A major disadvantage of the spontaneous Bragg scattering technique is
the limitation on the maximum number of photons (on the order of the
number of atoms in the cloud) that can actually be scattered off the
cloud before information on the internal state of the atoms is
destroyed by the spontaneous emission processes.

\section{Role of external
  magnetic fields} 

As most experiments on superfluidity in two-spin component Fermi atoms
require an external magnetic field $B_0$ to enhance atom-atom
interactions via the Feshbach resonance effect~\cite{ExpBCS_K,ExpBCS_Li}, it is
important to extend the discussion to this experimentally relevant case.
The geometrical issues are not as simple as discussed in
sec.\ref{sec:general}, so we shall limit ourselves to a few specific
examples 
of stimulated Bragg scattering processes and we shall postpone a more
complete analysis to a forthcoming publication.

Consider first a case where the magnetic field is weak enough not to
mix the $F_{g}=1/2$ ground state with other hyperfine components; this
approximation is expected to be valid provided the Zeeman energy
$\mu B_0$ ($\mu$ is the magnetic moment of the atom in the ground
state) is much smaller than the hyperfine splitting $\Delta_{HF}$. 
For the sake of simplicity, we assume that both Bragg beams are
polarized on the plane orthogonal to the magnetic field direction
(which is taken as the ${\hat z}$ axis).
In the circular $\pm$ basis, the Hamiltonian \eq{Ham_eff} is replaced by:
\begin{eqnarray}%
 \label{eq:light_matter_coupl_4}
  H^\pm_{\rm
  eff}&&=\int\!d\rr\,\sum_{ij=\pm}  f_{ij}\,{\hat I}_i(\rr)\,{\hat n}_j(\rr)= \\
=\frac{1}{2}\int\!d\rr\,\Big\{ &&\Big[ (f_{++}+f_{+-}){\hat I}_+(\rr) +
  (f_{-+}+f_{--}){\hat I}_-(\rr)
  \Big] {\hat n}(\rr) + \\
&+& \Big[ ((f_{++}-f_{+-}){\hat I}_+(\rr) +(f_{-+}-f_{--}){\hat I}_-(\rr)  \Big]\,{\hat S}_z(\rr) \Big\}
\end{eqnarray}%
where ${\hat I}_\pm(\rr)={\hat E}^\dagger_\pm(\rr)\,{\hat E}_\pm(\rr)$ are the
local light intensity operators in the $\sigma_\pm$ polarization state.  
The oscillator strenghts $f_{ij}$ for the $j=\pm$ Zeeman state
interacting with $\sigma_{i=\pm}$ polarized light have a non-trivial dependence
on the light frequency $\omega_L$ and the resonance frequencies
of the Zeeman-split transitions.
For the case of a transition to a $F_e=1/2$ state, no $\sigma_\pm$
transition can be driven on atoms in respectively the $\pm$ state, so 
$f_{++}=f_{--}=0$. On the other hand,
$f_{+-}=|d|^2/\hbar(\omega_L-\omega_{e,+1/2}+\omega_{g,-})$ and 
$f_{-+}=|d|^2/\hbar(\omega_L-\omega_{e,-1/2}+\omega_{g,+})$.
For the more complicate case of a transition to a $F_e=3/2$ state,
one has to take into account the relevant Clebsch-Gordan coefficients,
so that $f_{++}=|d|^2/\hbar(\omega_L-\omega_{e,+3/2}+\omega_{g,+})$, 
$f_{-+}= |d|^2/3\hbar(\omega_L-\omega_{e,-1/2}+\omega_{g,+})$ 
and analogous formulas hold for the $f_{+-}$ and $f_{--}$ oscillator
     strengths starting from the $g,-$ initial state.
Here, $\omega_{g,i=\pm}$ and
$\omega_{e,i=+3/2,+1/2,-1/2,-3/2}$ are the 
  energies of the Zeeman-split sublevels of the ground and excited states.

The analysis of sec.\ref{sec:stimulated} can then be easily repeated for this 
Hamiltonian: the main difference is in the perturbation
operator which now reads:
\begin{eqnarray}
{\hat A}&=&\frac{1}{2}\,\Big[\big(f_{++}+ f_{+-} \big)
  e^{-i\theta}+\big(f_{-+}+f_{--} \big) e^{i\theta}\Big]\,{\hat
  n}(\qq)+\\ &+&\frac{1}{2} \Big[ \big(f_{++}-f_{+-}\big)
  e^{-i\theta}+\big(f_{-+}-f_{--} \big) e^{i\theta}\Big]\,{\hat
  S}_z(\qq).
\end{eqnarray}
As previously, $\theta$ is the angle between the linear polarizations
of the two beams.
A non-vanishing coupling to the spin is present as soon as either
$f_{++}\neq f_{+-}$ or $f_{-+}\neq f_{--}$.

Of course, this framework is directly generalized to the case where the two
Zeeman sublevels of the ground state are replaced by a pair of
sublevels belonging to a higher angular momentum state, as in recent
experiments with $^{40}$K atoms trapped in their $F=9/2,
m_F=-9/2,-7/2$ states~\cite{ExpBCS_K}. The only difference is in the
Clebsch-Gordan coefficients appearing in the expression of the
oscillator strengths $f_{ij}$.
In $^{40}$K atoms, the hyperfine splitting $\Delta_{HF}\approx
870\,$MHz is in fact much larger than the Zeeman splitting at the
$B_0\simeq 200\,$G Feshbach resonance $\mu B_0\approx
280\,\textrm{MHz}\ll\Delta_{HF}$.
Further extension to the case where the two atomic states belong to different hyperfine
states is straightforward; this was indeed the case of the
state-selective imaging scheme used to study vortices in two-component Bose-Einstein
condensates~\cite{VorticesJILA}.

The case of $^6$Li is more complicated: around the $B_0\simeq 850\,$G
Feshbach resonance currently used in experiments on fermionic
superfluidity, the Zeeman splitting  
$\mu\,B_0\approx 1.2\,$GHz is substantially larger than the ground
state hyperfine splitting $\Delta_{HF}\simeq 225\,$MHz which makes the
level structure significantly different from the one discussed upto now.
We postpone a complete discussion of these issues to further
investigation, and we here limit ourselves to some general considerations.
In this regime, the atomic internal eigenstates are well approximated
by the ones of the Zeeman Hamiltonian, and the hyperfine
coupling is only a small perturbation. In particular, atoms are trapped in the two 
lowest-lying $\pm$ states in which the electron is always in the total
spin state $J=1/2$,$m_J=-1/2$ and the nucleus is in respectively the
$I=1$, $m_I=1,0$ spin state~\cite{Levels_Li}.
The fact that the electron is almost decoupled from the nucleus and is
in the $J=1/2$, $m_J=-1/2$ state implies that the oscillator strenghts
$f_{--}$ and $f_{-+}$ for $\sigma_-$-polarized light are very weak
(the matrix elements are a factor of the order $\eta=\Delta_{HF}/\mu
B_0$ smaller).
The coupling to the spin is mainly due to the difference $\Delta
f=f_{++}-f_{+-}$, and is provided by the hyperfine coupling in both the
ground and the excited states. This in fact splits the resonance
frequencies for the $\sigma_+$ transitions starting respectively from
the $\pm$ states. To make $\Delta f$ the largest, one has 
to choose a laser frequency close to resonance, but a compromise has
to be made with spontaneous emission.

\section{Conclusions}
In summary, we have shown how it is possible to take advantage of
the light polarization degrees of freedom to separately address the total
density and the spin density of a two-spin component atomic gas.
This can be exploited in polarization-sensitive Bragg scattering
techniques so as to measure both the density and the spin structure
factors of the gas and then to characterize the microscopic structure of
the gas in its different phases.
Stimulated Bragg scattering probes the response function of the gas to
the optical potential created by the interference pattern of the two
laser beams: one can act on either the total or the spin
density simply by changing the light polarizations.
The fluctuation-dissipation theorem then relates the absorbed energy
to the structure factor of our interest.
On the other hand, the density and spin structure factors can be
directly probed in a spontaneous Bragg scattering experiment: the
intensity of scattered light into a given polarization state is in
fact proportional to a combination of the density and the 
spin structure factors, whose weights depend on the relative angle
between the incident and scattered beam polarizations.

\vspace{0.5cm}

We are grateful to S. Stringari, C. Salomon, Y. Castin, and
S. Giorgini for continuous stimulating exchanges. Stimulating
discussions with R. Hulet and C. Regal are acknowledged. We
acknowledge hospitality at the Aspen Center for Physics and support
from the Ministero dell'Istruzione, dell'Universit\`a, e della Ricerca (MIUR).


\vspace{1.5cm}

\begin{thebibliography}{00}

\bibitem{SpinGeneral} 
H.J. Miesner, D.M. Stamper-Kurn, J. Stenger, S. Inouye, A.P. Chikkatur, and  W.
Ketterle,
Phys. Rev. Lett. {\bf 82}, 2228 (1999); 
A.E. Leanhardt, Y. Shin, D. Kielpinski, D.E. Pritchard, and W. Ketterle,
Phys. Rev. Lett. {\bf 90}, 140403 (2003); H. Schmaljohann, M. Erhard,
J. Kronjager, M. Kottke, S. van Staa, L. Cacciapuoti, J.J. Arlt, K. Bongs,
and K. Sengstock, Phys. Rev. Lett. {\bf 92}, 040402 (2004);
M.S. Chang, C.D. Hamley, M.D. Barrett, J.A. Sauer, K.M. Fortier,
W. Zhang, L. You, and M.S. Chapman,
Phys. Rev. Lett. {\bf 92}, 140403 (2004);  T. Kuwamoto, K. Araki, T. Eno, and T. Hirano,
Phys. Rev. A {\bf 69}, 063604 (2004).



\bibitem{VorticesJILA} M. R. Matthews, B. P. Anderson, P. C. Haljan,
  D. S. Hall, C. E. Wieman, and E. A. Cornell, Phys. Rev. Lett.{\bf 83},
  2498 (1999). 


\bibitem{SpinorStamper} J. M. Higbie, L. E. Sadler, S. Inouye,
  A. P. Chikkatur, S. R. Leslie, K. L. Moore, V. Savalli, and
  D. M. Stamper-Kurn, Phys. Rev. Lett. {\bf 95}, 050401 (2005) 



\bibitem{IC-EM} I. Carusotto and E. J. Mueller, J. Phys. B:
  At. Mol. Opt. Phys. {\bf 37}, S115 (2004)

\bibitem{ExpBCS_K} C. A. Regal, M. Greiner, and D. S. Jin,
Phys. Rev. Lett. {\bf 92}, 040403 (2004).

\bibitem{ExpBCS_Li} M. W. Zwierlein {\it et al.},
Phys. Rev. Lett. {\bf 92}, 120403 (2004); 
M. Bartenstein, A. Altmeyer, S. Riedl, S. Jochim, C. Chin, J. Hecker Denschlag, and R. Grimm
Phys. Rev. Lett. {\bf 92}, 203201 (2004); T. Bourdel, L. Khaykovich,
J. Cubizolles, J. Zhang, F. Chevy, M. Teichmann, L. Tarruell,
S. J. J. M. F. Kokkelmans, and C. Salomon 
Phys. Rev. Lett. 93, 050401 (2004).

\bibitem{TheoryBCS} A. J. Leggett, J. Phys. (Paris) C {\bf 7}, 19 (1980);
P. Nozi\`eres, S. Schmitt-Rink, J. Low Temp. Phys. {\bf 59}, 195 (1985);
J. R. Engelbrecht, M. Randeria, and C. A. R. S\'a de Melo, Phys.
Rev. B {\bf 55}, 15153 (1997).



\bibitem{SupercondBooks} J. R. Schrieffer, {\em  Theory of
  superconductivity}, Addison-Wesley, Redwood City, 1988; D. R. Tilley
  and J. Tilley, {\em Superfluidity and superconductivity}, Hilger,
  Bristol, 1990.

\bibitem{DensityCorrelationsExp} M. Greiner, C. A. Regal, C. Ticknor,
  J. L. Bohn, and D. S. Jin,  Phys. Rev. Lett. {\bf 92}, 150405
  (2004); G. B. Partridge, K. E. Strecker, R. I. Kamar, M. W. Jack, and R. G. Hulet
Phys. Rev. Lett. {\bf 95}, 020404 (2005).



\bibitem{Bragg} D. M. Stamper-Kurn, A. P. Chikkatur, A. G\"orlitz,
  S. Inouye, S. Gupta, D. E. Pritchard, and K. Ketterle,
  Phys. Rev. Lett. {\bf 83}, 2876 (1999). 

\bibitem{CCT} C. Cohen-Tannoudji and J. Dupont-Roc, Phys. Rev. A {\bf
  5}, 968 (1972).

\bibitem{Texas} 
A. M. Dudarev, R. B. Diener, I. Carusotto, and Q. Niu,
  Phys. Rev. Lett. {\bf 92}, 153005 (2004) 



\bibitem{Yvan} A. Minguzzi, G. Ferrari, and Y. Castin, Eur. Phys. J. D
  {\bf 17}, 49  (2001)

\bibitem{Buchler} H. P. B\"uchler, P. Zoller, and W. Zwerger
Phys. Rev. Lett. {\bf 93}, 080401 (2004)  



\bibitem{Lattice} J. Dalibard and C. Cohen-Tannoudji,
  J. Opt. Soc. Am. B {\bf 6}, 2023 (1989); O. Mandel, M.
  Greiner, A. Widera, T. Rom, T. W. H\"ansch, and I. Bloch,
  Phys. Rev. Lett. {\bf 91}, 010407 (2003).

\bibitem{AndersonPhonon} P. W. Anderson, Phys. Rev. {\bf 112}, 1900
  (1958).

\bibitem{LinearResponse} 
L. D. Landau, E. M. Lifshitz, and L. P. Pitaevskii,
  {\em Statistical Physics}, Vols.1 and 2, Pergamon Press, Oxford,
  1980; L.-P. L\'evy, {\em Magn\'etisme et
  supraconductivit\'e}, Inter\'Editions / CNRS \'Editions, Paris,
  1997, chapter 8.

\bibitem{f-sum} L.P. Pitaevskii and S. Stringari, {\sl Bose-Einstein Condensation},
Clarendon Press Oxford (2003).

\bibitem{A-M} N. W. Ashcroft, and N. D. Mermin, {\em Solid State
  Physics}, Sauders College Publishing, 1976.


\bibitem{Levels_Li} M. E. Gehm, PhD thesis, Duke University, 2003; 
  M. E. Gehm, {\em Properties of $^6Li$}, {\tt
  http://www.phy.duke.edu/research/photon/qoptics/techdocs/pdf/PropertiesOfLi.pdf} 


\end{thebibliography}
\end{document}